\documentclass[useAMS,usenatbib]{mn2e}
\usepackage{graphicx,amssymb}
\bibpunct[,]{}{}{;}{a}{}{,}

\usepackage{amssymb}

\title[Spectroscopic analysis of HD\,207561] {A spectroscopic analysis of the chemically peculiar star HD\,207561\thanks{The present work is based on the analysis of data collected with the Russian 6-m telescope BTA operated by the Special Astrophysical Observatory of the Russian Academy of Sciences (SAO RAS)}}

\author[S. Joshi et al.]{S. Joshi$^{1}$\thanks{E-mail: santosh@aries.res.in},
E.~Semenko$^{2}$, P. Martinez$^{3}$, M. Sachkov$^4$, Y. C. Joshi$^{1}$, S. Seetha$^{5}$, 
\newauthor{N. K. Chakradhari$^{6}$, D. L. Mary$^7$, V. Girish$^5$, B. N. Ashoka$^{5}$ }\\ 
\\
$^{1}$Aryabhatta Research Institute of Observational Sciences, Manora peak, Nainital-263129, India	\\
$^{2}$Special Astrophysical Observatory RAS, Nizhny Arkhyz, Karachai-Cherkassian Republic, 369167, Russia\\
$^{3}$South African Astronomical Observatory, PO Box 9, Observatory 7935, Cape Town, South Africa	\\
$^{4}$Institute of Astronomy, Russian Academy of Sciences, Pyatnitskaya 48, 119017 Moscow, Russia \\
$^{5}$ISRO Satellite Center, Airport Road, Bangalore-560 017, India \\
$^{6}$School of Studies in Physics and Astrophysics, Pt. Ravishankar Shukla University, Raipur, India \\ 
$^{7}$Laboratoire Lagrange, UMR 7293, Universit\'e de Nice Sophia-Antipolis,    CNRS, Observatoire de la  C\^{o}te d'Azur}

\begin{document}

\date{Accepted Received }

\pagerange{\pageref{firstpage}--\pageref{lastpage}} \pubyear{2011}

\maketitle
\begin{abstract}

In this paper we present a high-resolution spectroscopic analysis of the chemically peculiar star HD\,207561. 
 During a survey programme to search for new roAp stars in the Northern hemisphere, Joshi et al. (\cite{josh06}) observed significant photometric variability on two consecutive nights in the year 2000. The amplitude spectra of the light curves obtained on these two nights showed oscillations with a frequency of 2.79 mHz [P$\sim$6-min]. However,  subsequent follow-up observations could not confirm any rapid variability. In order to determine the spectroscopic nature of HD\,207561, high-resolution spectroscopic and spectro-polarimetric observations were carried out. A reasonable fit of the calculated $H_\beta$ line profile to the observed one yields the effective temperature ($T_\mathrm{eff}$) and surface gravity ($\log g$) as 7300 K and 3.7 dex, respectively. The derived projected rotational velocity ($v\sin i$) for HD\,207561 is  74 km{\,s$^{-1}$} indicative of 
a relatively fast rotator. The position of  HD\,207561 in the H-R diagram implies that this is slightly evolved from the main-sequence and located well within the $\delta$-Scuti instability strip. The abundance analysis indicates the star has slight under-abundances of Ca and
Sc and mild over-abundances of iron-peak elements. The spectro-polarimetric study of HD\,207561 shows that  the effective magnetic field is within the observational error of 100 gauss (G). The spectroscopic analysis revealed that the star has most of the characteristics similar to an Am star, rather than an Ap star, and that it lies in the  $\delta$-Scuti instability strip; hence roAp pulsations are not expected in HD\,207561, but low-overtone modes might be excited.

\end{abstract}


\section{Introduction} \label{intr1}
In contrast to normal stars, the chemically peculiar (CP) stars are identified by the presence of abnormally strong and/or weak absorption lines of certain elements in their spectra. The two sub-groups of CP stars namely Am (metallic-lined) and Ap (A-peculiar) are important asterosesmic tools because some members of these classes show multi-periodic pulsational variability. The Am stars which are non-magnetic in nature are mostly found in short period binary systems with orbital periods between 1 to 10 d, causing synchronous rotation with {$v \sin i$ $\leq$ 120 km$s^{-1}$} (Abt \cite {abt09}). Now many Am stars are known to exhibit low-amplitude, multi-periodic variability with period range 0.25 to 7 hrs i.e. similar to the $\delta$-Scuti variables (Joshi et al. \cite{josh03}, \cite{josh06}, \cite{josh09}; Smalley et al. \cite{smal11}). The Ap stars showing magnetic field strengths of the order of kilogauss (kG) exhibit variability of periods in the range 5 to 21 min­ with Johnson B amplitudes  $<$ 8 mmag and spectroscopic radial velocity variations of 0.05 to {5 km\,s$^{-1}$} (Kurtz, Elkin \& Mathys \cite{kurt06}), are known as rapidly oscillating Ap (roAp) stars.  There is no clear correlation between photometric and spectroscopic pulsation amplitude in roAp stars but  the pulsation signal can be detected in spectroscopic observations  with no detection in the photometry (Hatzes \& Mkrtichian (\cite{hatz04})~-- $\beta$\,CrB; Kochukhov et al. (\cite{kochu09}) -- HD\,75445). The roAp stars possess strong abundance of rare earth elements (REE) and almost all of them show ionization disequilibrium for Pr and Nd. Currently, about 46 such candidates are known (Joshi et al. \cite{josh12}). The light curves of many roAp stars show a double modulation. Such modulation in these stars is explained by the oblique pulsator model (Kurtz \cite{kurt82}). According to this, the pulsation axis is aligned with the magnetic axis, which is itself inclined to the rotation axis. As the star rotates the observer views the pulsation modes from an aspect that varies with rotation. Treating the combined effects of rotation and magnetic field on the pulsation modes, Bigot \& Dziembowski (\cite{bigo02}) proposed the revised pulsator model where the pulsation axis lies in between the rotation and magnetic axis. In the framework of the modified oblique pulsation model,  Bigot \& Kurtz (\cite{bigo11}) presented a theoretical and analytical study of the light curves associated with dipole ($l$ = 1) pulsations of roAp stars. Excitation of pulsations in pulsating Am and roAp stars is governed by the $\kappa$-mechanism operating in the partial hydrogen ionization zone (Balmforth et al. \cite{balm01}; Cunha \cite{cunh02}; Vauclair \& Theado \cite{vauc04}). There are a few Ap stars which show pulsation periods in the range of $\delta$-Scuti stars (Kurtz et al. \cite{kurt08}; Gonzalez et al. \cite{gonz08}) but not a single Am star is  known to date where rapid oscillations are observed. Therefore, the pulsations in Am and Ap stars are important astrophysical tools to study the complex relationship between stellar pulsation and magnetic field  in the presence of atmospheric abundance anomalies. 

 The ``Nainital-Cape'' survey was initiated in 1999 between Aryabhatta Research Institute of Observational Sciences (ARIES), Nainital and the South African Astronomical Observatory (SAAO), South Africa. The main aim of this survey project was to search and study for such Ap and Am stars which are pulsationally unstable. The major results of  this survey were published by Martinez et al. (\cite{mart01}); Joshi et al. (\cite{josh06},~\cite{josh09}). 

Based on published Str\"omgren photometric indices similar to those of the then known roAp stars, we selected HD\,207561 as a target star  in the year 2000 (JD\,2451832) to search for rapid photometric variability on a time scale of 6 to 16 min. The time-series series photoelectric photometry of this star obtained on two 
consecutive nights showed an apparent photometric variability with a period $\sim$ 6 min (Joshi et al. \cite{josh06}). However, subsequent follow-up observations over a period of eight years did not confirm any such  rapid variability. Therefore to establish the spectral classification and magnetic nature of HD\,207561 we subjected it for high-resolution spectroscopic and spectro-polarimetric observations. In this paper we present the spectroscopic analysis of this star. The paper is organized as follows: In Sec. \ref{intr2} we discuss the various astrophysical parameters calculated from the standard relations. The observations and data reduction procedure is described in Section \ref{obs}. The spectroscopic data analysis is presented in Sec. \ref{spectroana}. Finally, we discuss and concluded our results in Sec. \ref{disc}.

\section{HD\,207561} \label{intr2}

The star HD\,207561 ($\alpha_{2000}$ = 21 48 16; $\delta_{2000}$ = 54 23 15, V = 7.84 mag) is classified as F0 III/F0 IV (Olsen \cite{olse83}). The Str\"{o}mgren indices of this star are $b-y$ = 0.142, $m_1$ = 0.220, $c_1$ = 0.820 and $\beta$ = 2.825 (Hauck \& Mermilliod \cite{hauc98}). From the calibrations for A-type stars given by Crawford(\cite{craw79}), we derive $E(b-y) = 0.018$, $\delta m_0$ = $-$0.022 and $\delta c_0$ = $-$0.003. The $\delta m_0 $ and $\delta c_0$ indices indicate strong line blocking in the $u$ and $v$ filters and are typical of strongly peculiar Am and Ap spectra. Cowley \& Cowley (\cite{cowl65}); Bertaud \& Floquet (\cite{bert74}); Nicolet (\cite{nico83}) classified this as a marginal\footnote{The marginal Am stars  are those for which  the difference between the spectral type determined from the CaII K line and from the metal lines is less than 5 spectral subclasses. If the difference is more than  5 spectral subclasses then they are refereed as classical Am stars.} Am star. 

 An apparent pulsation period of 6 min observed on two consecutive nights made HD\,207561 an interesting object for the further study. The knowledge of the stellar parameters allows us to clarify the general view on the nature of the studied star. Therefore, we estimated the astrophysical parameters for HD\,207561 using the Str\"omgren and Geneva indices taken from the SIMBAD database.

 The galactic coordinates of HD\,207561 ($l=98.2598^{\circ}$ and $b=0.5360^{\circ}$) imply its location near the galactic plane. Taking into account the trigonometric parallax of HD\,207561 $\pi=8.57\pm0.56$ mas (van Leeuwen  \cite{vanl07}) a non-zero interstellar extinction is expected. This value can be determined using the three independent methods. The calibration of Moon \& Dworetsky (\cite{moon85}) using Str\"omgren indices gives $E(b-y)=0.018$ and hence $E(B-V)=0.026$. The equivalent width of the interstellar Na D1 line, available in our Echelle spectrum gives $E(B-V)=0.050$ (Munari \& Zwitter \cite{muna97}). The value of $E(B-V)$ derived from reddening maps of Lucke (\cite{luck78}) is 0.070 is thought to be overestimated. Due to the higher accuracy of the first two methods, for the present study we adopted an average value $E(B-V)$= 0.038. This corresponds to an interstellar extinction $A_{V}=0.118$ mag.

\begin{figure}
\begin{center}
\includegraphics[width=7.0cm,height=8.5cm,angle=-90]{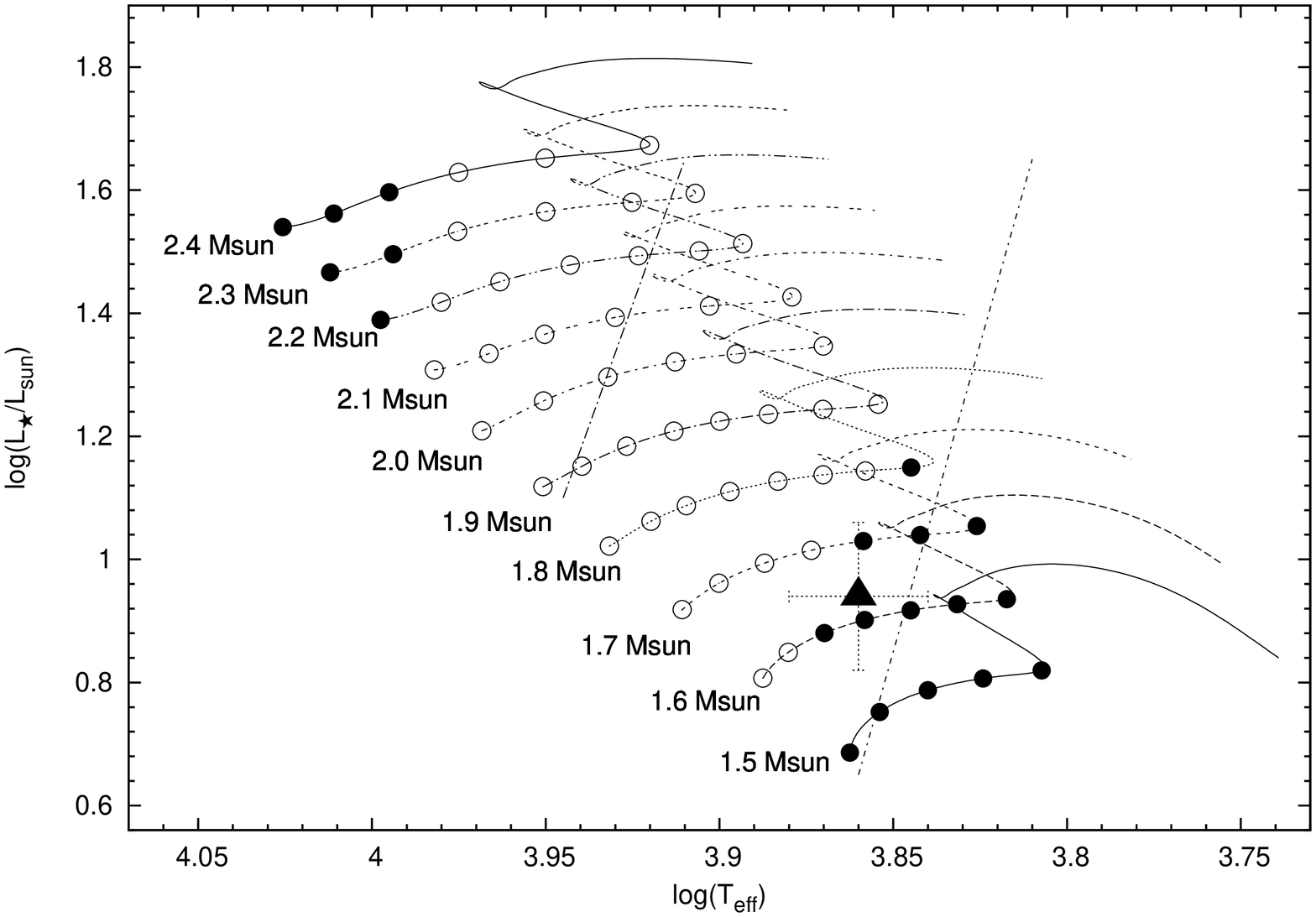}
\caption{The location of HD\,207561 in the H-R diagram is shown by a triangle. The filled circles and open circles indicate the locations of  pulsationally stable and unstable modes for roAp stars (Cunha \citet{cunh02}), respectively. Two vertical lines show the boundary of the $\delta$-Scuti
instability strip (Turcotte \citet{turc00}). The star HD\,207561 resides well within the main sequence region of the $\delta$-Scuti instability strip.}
\label{HR}
\end{center}
\end{figure}

The effective temperature of HD\,207561 could be determined from the available photometric data. Based on the Str\"omgren system two types of calibrations were implemented through the TempLogg programme (St\"utz et al. \cite{stut02}). From the calibration of Moon \& Dworetsky (\cite{moon85}) the derived value of effective temperature,  gravity, absolute magnitude and metallicity are $T_{eff}=7908\pm193$\,K, $\log g=4.17\pm0.10$ dex, $M_{V}$=2.52 mag and $[M/H]=0.22$, respectively. The calibration of Napiwotzki, Schoenberner \& Wenske (\cite{napi93}) gives the value of $T_{eff}$ = $7650\pm150$ K and $\log g=4.07\pm0.10$ dex. The photometric data in Geneva system were also used to derived these basic parameters. Applying the calibration of Kunzli et al. (\cite{kunz97}) on the Geneva system, the derived parameters $T_{eff}$, $\log g$ and $[M/H]$ are $7884\pm80$\,K, $4.59\pm0.06$ dex and 0.19 respectively. We note that these calibrations are slightly less reliable for stars with peculiar abundances and stratified atmosphere than for normal stars. However, Smalley \& Dworestsky (\cite{smal93}) concluded that values for T$_{eff}$ and $\log g$ determined from photometry are extremely reliable and not significantly affected by metallicity. Therefore we can assume that the average photometric value of effective temperature and surface gravity are  $7814$\,K and $\log g=4.07$, respectively. The spectroscopic determination of these parameters is described in Sec. \ref{stelpara}.

The bolometric corrections($BC$) from interpolation in the tables by Flower (\cite{flow96}) is estimated as $BC=0.031$ for $T_{eff}= 7815\,K$  and $BC=0.035$ for $T_{eff}=7300$\,K. The lower value of $T_{eff}$ corresponds to the temperature estimated from fitting of hydrogen line H$_{\beta}$.  The revised HIPPARCOS parallax $\pi=8.57\pm0.56$ mas (van Leeuwen  \cite{vanl07}) and an interstellar extinction $A_{V}=0.118$ mag resulting the absolute magnitude $M_{V}$=2.39 mag. The luminosity parameter $\log (L_{*}/L_{\odot})$ for HD\,207561 is 0.91 mag. Using the standard relation the calculated mass and radius of the star are $M=1.65\,M_{\odot}$ and $R=1.59\,R_{\odot}$, respectively. The lower temperature ($T_{eff}=7300$ K) yields the larger value for the radius ($R=1.82\,R_{\odot}$) and smaller value for the surface gravity ($\log g=4.18$). 
These basic parameters  are well within the range of the $\delta$-Sct and roAp stars. The location of the star in the H-R diagram is shown in Fig. \ref{HR}. The open circles in the figure indicate the predicted frequencies for pulsationally unstable rapid oscillation modes while the pulsationally stable modes are shown by filled circles (Cunha \cite{cunh02}).  The evolutionary tracks\footnote{http://www.phys.au.dk/$\sim$jcd/emdl94/eff$\_$v6} of masses ranging from 1.5 to 2.4 $M_\odot$ are also plotted (Christensen-Dalsgaard \cite{chri93}). From Fig. \ref{HR},  it is evident that HD\,207561 is slightly evolved from the main sequence and lies well within the $\delta$-Scuti instability strip. Therefore, HD\,207561 might exhibit pulsations similar to $\delta$-Scuti type variables.

\section{Observations and data reduction} \label{obs}
Due to its potential asteroseismic importance, HD\,207561 was observed in both the spectroscopic and spectro-polarimetric modes. The following sub-sections briefly summarize the details of the observation and data reduction procedure.

\begin{figure}
\begin{center}
\includegraphics[width=90mm]{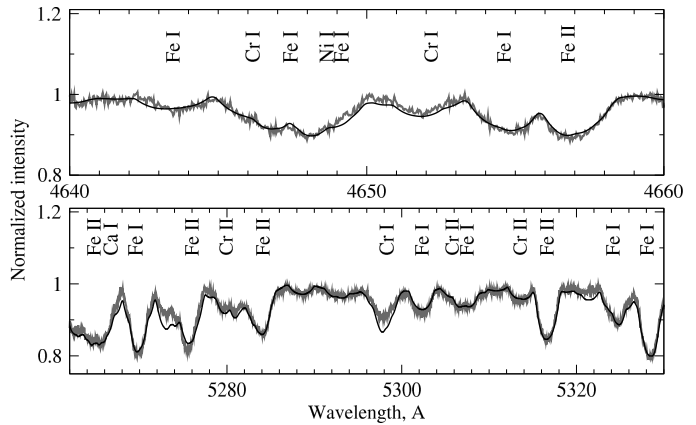}
\caption{Two sections of Echelle spectrum of HD\,207561 with identified lines.}
\label{conti}
\end{center}
\end{figure}

\subsection{High-resolution Spectroscopy}\label{spectrosc}

High-resolution spectroscopic observations were carried out from the 6.0-m BTA telescope at the Special Astrophysical Observatory, Nizhny Arkhyz, Russia (SAO). The first spectroscopic data were obtained on 17 November 2008 ($\mathrm{JD} 2454788.104$) with a high-resolution Nasmyth Echelle Spectrometer~(NES)~(Panchuk  et al. \cite{panc09}) installed on a Nasmyth platform. A single ThAr reference spectrum, obtained immediately
after the target star spectra  were acquired for wavelength calibration purposes.  The data reduction procedure consists of bias subtraction, flat-fielding, stray light correction and recognizing of spectral orders with further extraction of them.  All listed procedures are implemented in a set of IDL routines called REDUCE~(Piskunov \& Valenti \cite{pisk02}), and have been executed in a semi-automatic mode. The final one-dimensional spectrum covers the wavelength range from
4460 to 5930 \AA, with a signal-to-noise ratio of about 150. The mean
resolving power of spectrometer R = 39000 at that observational setting was determined from the Th-Ar lines. The continuum normalization was done with the IRAF\footnote{IRAF is distributed by the National Optical Astronomy Observatories, which are operated by the Association of Universities for Research in Astronomy, Inc., under cooperative agreement with the National Science Foundation.} task {\sc continuum}. Fig. \ref{conti} shows  two parts of the normalized spectrum where the various elements are listed. The broadening of the various lines show the clear indication of fast rotation in HD\,207561.

\begin{figure}
\begin{center}
\includegraphics[width=90mm]{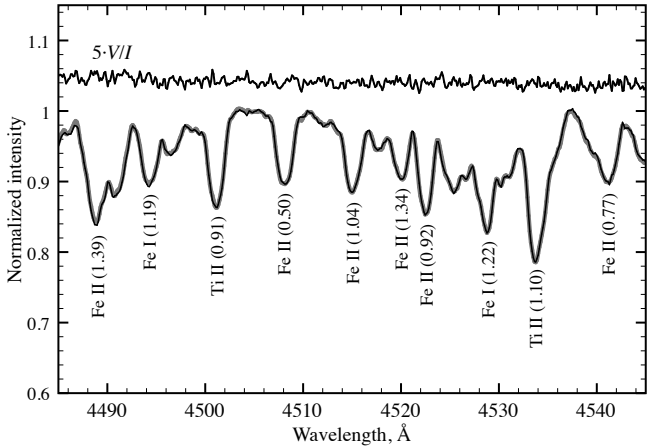}
\caption{The polarized spectrum of HD\,207561. A scaled and shifted up Stockes' V parameter V/I= $I_1$ - $I_2$ / $I_1$ + $I_2$, where $I_1$ and $I_2$ are the intensity of right and left circularly polarized spectrum, respectively, is also plotted. Gray and black profiles are circularly polarized stellar spectra. For individual lines designation of an ion and Land\'{e} factor is written.}
\label{pol}
\end{center}
\end{figure}

\subsection{Spectro-polarimetry}\label{spectropol}

A strong global-scale magnetic field is an intrinsic property of most of roAp stars. To test whether HD207561 might be an Ap star, and hence exhibit rapid oscillations, as potentially observed by Joshi et al (2006), we searched for evidence of a magnetic field in HD\,207561. The long-term usage of the above mentioned telescope together with the Main stellar spectrograph (MSS), equipped with circular polarization analyzer, confirm its applicability for  such kind of searches~(Kudryavtsev et al. \cite{kudr06}). MSS of BTA is a long-slit spectrograph equipped with circular polarization analyzer that is  combined with an image slicer~(Chountonov \cite{chou04}). The analyzer has a rotatable quarter-wave plate that is able to take two fixed positions corresponding to the angles 0 and 90$^{\circ}$ relative to the birefringence crystal. The spectra were taken with a EEV CCD detector ($4600\times2048$ pixels size). The mean resolving power  of spectrograph was about 15,000. Each scientific exposure actually consists of two subsequent frames when the quarter-wave plate had opposite orientation. We obtained three Zeeman spectra, one on 21 May 2011 (JD2455702) and other two on 22 July 2011 (JD2455764). The  time interval between first two spectra was enough to detect magnetic field in the case of probable long-term variability. The log of spectro-polarimetric observations is given in the Table~\ref{speclog}. The data reduction of spectro-polarimetric data was performed by means of a set of routines {\sc Zeeman} written in SAO by D.~O. Kudryavtsev for the ESO MIDAS system. The sequence of reduction stages is similar to the reduction scheme for echelle spectra except for some differences caused by a long-slit character of the data.  For the illustrational purposes we show the polarized spectrum of HD\,207561 in Fig. \ref{pol}.  The value of Stokes parameter is rather small, hence we scaled it by a factor 5 and shift upward. SNR of the spectra was high and due to zero magnetic field, polarized spectra are practically not resolved.

\begin{table}
\caption{Spectro-polarimetric observational log of HD\,207561.}
\label{speclog}
\begin{center}
\begin{tabular}{c c c}
\hline
JD & Wavelength range & SNR \\
2450000+&  \AA\ & \\
\hline
5702.506  & $4426-4978$ & 230 \\
5764.342  & $4420-4974$ & 250 \\
5764.508  & $4420-4974$ & 200 \\
\hline
\end{tabular}
\end{center}
\end{table}

\section{Spectroscopic Analysis}\label{spectroana}

\subsection{The Stellar Parameters}\label{stelpara}

\begin{figure}
\centering\includegraphics[width=82mm]{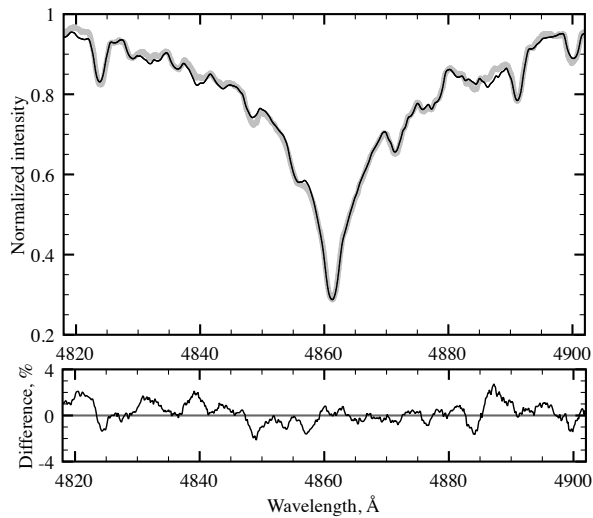}
\caption{The comparison of observed profile of $H_\beta$ line~(gray line) with synthetic one~(black line) computed for effective temperature 7300 K, surface gravity 3.7 dex, and metallicity +0.2. The bottom panel show the residual of the fit.}
\label{hydr}
\end{figure}

In order to find the effective temperature and surface gravity of HD\,207561, the fitting of synthetic profile of hydrogen line H$_{\beta}$ to the observed spectrum was also used, additionally to the photometric methods listed in Sec. 2.  The {\sc Synth3} programme code writen by O.~Kochukhov (Kochukhov \cite{kochu07}) was used to compute the grid of synthetic spectra in the range 4780--4930\,\AA. This code compute the synthetic spectra in the LTE regime.  The Kurucz's model of stellar atmosphere with convection were computed with {\sc Atlas9} code (Kurucz \cite{kuru93}). The list of lines from Vienna Atomic Line Database~(Piskunov et al. \cite{pisk95}; Kupka et al. \cite{kupk99}) acted as input for the further spectrum synthesis. The computed spectra were then compared with an observed spectrum of HD 207561. The best fit was achieved with the parameters set: $T_{eff}=7300\pm 250$ K, $\log g=3.7\pm0.1$ dex and $[M/H]=+0.2$. The upper-panel of Fig. \ref{hydr} shows the best fit between the observed and synthetic spectrum and the lower panel shows the residuals of the fit.

\subsection{Chemical Abundances}

The presence of high over-abundances of rare earth elements
is one of the known characteristics of the roAp stars. The difference by more than 1 dex between abundances measured from first two ionized states of Nd, Pr and other REEs is an another important property of these stars. It is believed that doubly ionized ions of REEs are formed in higher layers of a stellar atmosphere (Ryabchikova et al. \cite{ryab04}; Kurtz, Elkin \& Mathys \cite{kurt07}).

 The projected rotational velocity of HD\,207561 is $v\sin i=74\pm5$ km\,s$^{-1}$  and indicates a short rotation period. Our analysis of the available spectra showed the clear presence of lines of Fe, Cr, Mg and some other elements are present in the blends. Comparing the observed spectra with those calculated with the {\sc Synth3} code, we determined the abundances of 14 elements. Assuming the micro-turbulence velocity $\xi_{micro}=2$ km\,s$^{-1}$, we have been used the method of synthetic spectra to fit the observed profiles. Our choice of $\xi_{micro}$ is based on the unknown nature of the star. $\xi_{micro}=2$ km\,s$^{-1}$ is more typical for chemically peculiar Ap stars, while the metallic-line stars usually have the higher velocities. However,  $\xi_{micro}=2$ km\,s$^{-1}$ is a good approximation for many of Am stars.

\begin{figure}
\begin{center}
\includegraphics[width=90mm]{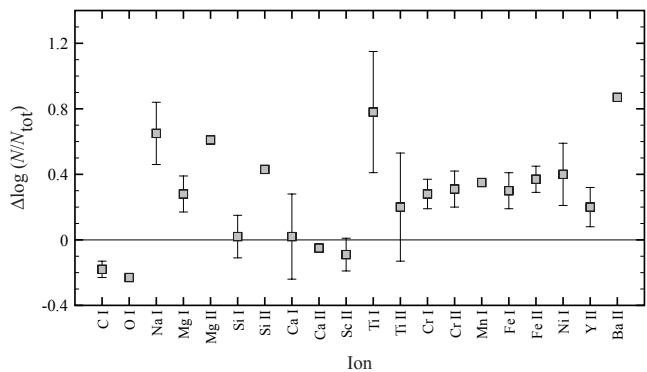}
\caption{Derived atmospheric abundances for 14 elements relative to the Sun (Asplund, Grevesse \& Sauval \citet{aspl05}).}
\label{abundfig}
\end{center}
\end{figure}

The results of measurements are presented in Table \ref{abundtab} and shown graphically in Fig. \ref{abundfig}. These results are based on the assumption of the effective temperature and surface gravity values as they were obtained from spectroscopy. It can be seen from Table \ref{abundtab} that C, O, Ca and Sc shows a mild deficiency while the other elements have excess of abundances, compared to solar values. Such an abundance pattern is typical for metallic-line stars. However, it is not so clear if we take the atmospheric parameters derived from photometry. In this case the most of individual abundances will increase by about 0.2-0.3 dex and all Am star attributes will disappear. In available spectra we cannot find any significant lines of REEs but the modeling of a few Nd lines from the structure of blends confirm near-solar abundance of this element. So, we can conclude that HD 207561 is most probably an Am star with mild chemical peculiarities or normal A star. The absence of significant anomalies typical for magnetic CP stars reject the assumption about its magnetic nature.

\subsection{Magnetic Field Measurements}\label{mag}

Our measurements of the effective longitudinal magnetic field of the star were made by means of the classical method where a value of $B_\mathrm{e}$ is determined from the relation
\begin{equation}
B_\mathrm{e} = \frac{\Delta\lambda}{9.34\times10^{-13}\,\times\lambda_{0}^{2}\,\times\overline{g}}\,\,\mathrm{[G]}.
\end{equation}
In this equation $\Delta\lambda$ is the difference between the position of the same line in right and left circularly polarized spectra, $\lambda_{0}$ is a wavelength of a line in intensity spectrum and $\overline{g}$ is the mean value of Lande factor. We have used the value of $\overline{g}$ as 1.23, approximately equal to averaged value of individual Lande factors of lines in selected spectral range (Romanyuk \cite{roma84}). Some of the authors  used slightly different value (1.21 or similar)  for the same purposes that is not crucial for the final result.
Individual positions of separate lines in polarized spectra were measured using the center-of-gravity method. Results of the magnetic measurements are presented in  Table~\ref{tab:zeeman}.

\begin{table}
\caption{Results of the measurements of effective longitudinal magnetic field of HD 207561, $n$ is the number of lines used for the analysis.}
\label{tab:zeeman}
\begin{center}
\begin{tabular}{c c c}
\hline
JD  &  $B_\mathrm{e}\pm\sigma$ & $n$ \\
2450000+&  G & \\
\hline
5702.506  & $-62\phantom{\pm}91$ & 65 \\
5764.342  & $70\phantom{\pm}93$  & 57 \\
5764.508  & $45\phantom{\pm}124$ & 57 \\
\hline
\end{tabular}
\end{center}
\end{table}
 
On the night of 21 May 2011 we observed Arcturus and $\alpha^2$ CVn as zero-field and magnetic standard, respectively. The respective derived longitudinal field for these stars are  205$\pm$50 G (n=249 lines) and 950$\pm$68 G (n=133 lines).
The value of magnetic field in $\alpha^2$ CVn is corrected for instrumental non-zero polarization measured from Arturus. Within the error our measured value is close to the predicted value of $B_e$ for $\alpha^2$ CVn i.e. about 900-950 G (Wade et al. \cite{wade00}). Further, two standard stars HD\,158974 and $\beta$ CrB 
were also observed on night 22 July 2011. The measured longitudinal field of the first, zero-field standard, star is $0\pm50$ G ($n=305$ lines). Polarized spectra of the well-known magnetic star $\beta$ CrB shows the Zeeman shift corresponding to the longitudinal magnetic field $B_{e}=-170\pm50$ G ($n=240$ lines). For this star at rotational phase 0.27, Wade et al. (\cite{wade00}) predicts the field of about $-100$ G. The measured longitudinal
 magnetic field (corrected for instrumental polarization) in HD\,207561 is given in Table~\ref{tab:zeeman}. From the Table we can conclude that the star HD\,207561 does not exhibit any magnetic property.

\begin{figure}
\begin{center}
\includegraphics[width=9cm,height=10cm]{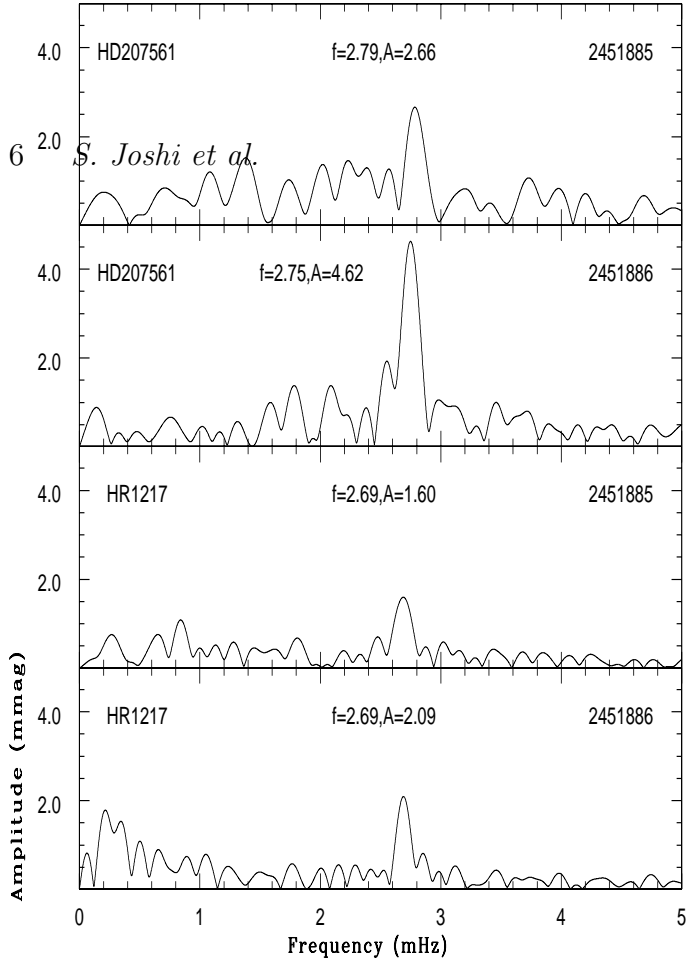}
\caption{The $B$-band amplitude spectra of HD\,207561 and HR\,1217 obtained on 
two consecutive nights. HR\,1217 is a well known multi-periodic roAp star and has one pulsational frequency at 2.79 mHz, the same frequency observed on the two time-series data sets of HD\,207561. The frequency spectra of both the stars show a prominent peak with night-to-night amplitude  modulation.}
\label{lc-ps}
\end{center}
\end{figure} 

\section{Discussions and Conclusions} \label{disc}

The transient photometric variability observed on two consecutive observing runs placed HD\,207561 in the category of potentially interesting variable stars. For the ``Nainital-Cape'' survey we observed each candidate star for a duration of 1--2 hr that is sufficient to reveal the roAp-like oscillations in any single photometric night (Martinez et al. \cite{mart01}). Therefore, 3 to 4 candidate stars could be monitored  on a particular night. On the two nights when the photometric variability of 6 min was observed in HD\,207561, a well known roAp star HR\,1217 was also monitored under a multi-site campaign (Kurtz et al. \cite{kurt02}).  
The top two panels of Fig. \ref{lc-ps} show the amplitude spectra of HD\,207561 while the bottom two panels show the amplitude spectra of HR\,1217 on the same nights. The noise levels in the amplitude spectra of both stars are comparable to each other. HR\,1217 is a multi-periodic roAp star where seven frequencies have been detected close to 2.69 mHz and one of them correspond to 2.79 mHZ. On those two particular nights, HD207561 appeared to show similar rapid oscillations.

\begin{figure*}
\begin{center}
\includegraphics[width=18cm,height=18cm]{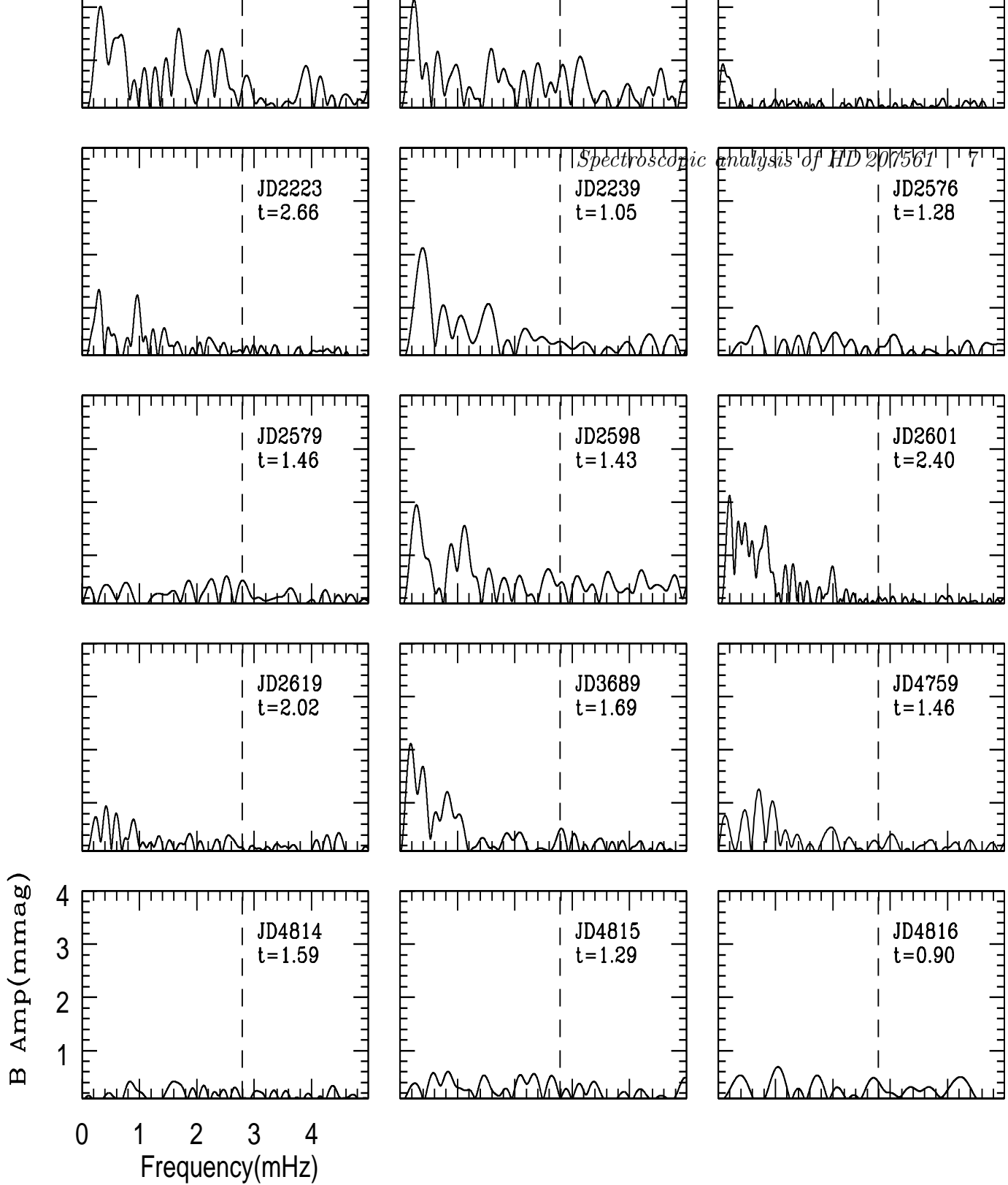}
\caption{Samples of the pre-whitened amplitude spectra of HD\,207561. The vertical dashed line indicate the frequency where photometric variability of $\sim$ 6 min was detected on two consecutive nights of year 2000.  Each
panel contains the Fourier transform of an individual light curve, covering a frequency range of $0$ to $5$\,mHz, and an amplitude range of 0 to 4 mmag. The date of the observation in Julian date (JD 245000+) and the duration of observations (t in hours) are mentioned in each panel.}
\label{prewhite}
\end{center}
\end{figure*}

In an attempt to confirm the presence of rapid oscillations in HD207561, we observed the star again on many nights between 2000 and 2008. Fig. \ref{prewhite} shows the samples of pre-whitened amplitude spectra of HD\,207561. In none of these observations we see any indication of rapid oscillations near to frequency 2.79 mHz. The observations were carried out with the same instrument, a three-channel photo-electric photometer used in the ``Nainital-Cape'' survey (Ashoka et al. \cite{asho01}). Generally, the observations were done in single-channel mode but on some occasions a nearby faint comparison star was also observed in the second channel, to test whether this might have been an instrumental effect in both channels. However, the amplitude of variation in the second channel was less than 2 mmag and no coherent pulsations were seen in the data of the comparison star (Girish \cite{giri05}). In order to test for a positional dependence of the instrumental noise we pointed the telescope at declination of HD\,207561 and the observations were acquired at same hour angle range but with the photometer dark slide shut. In this test we did not notice any periodic variations in the dark counts. The electronics used in the photometer and alignment of its optical elements were also checked and did not find any irregularities in the instruments.

\begin{table}
\begin{center}
\caption{ Abundances of HD\,207561 as determined for the case of  $T_{eff}$=7300 K and log g=3.7 dex is listed in the second column. The third column is the difference in abundances between the photometric and the spectroscopic sets of parameters. For the comparison the abundances of the solar atmosphere is also presented (Asplund, Grevesse \& Sauval \citet{aspl05}).} 

\begin{tabular}{l c c|c}
\hline
Ion  &  HD\,207561  & HD\,207561  &  Sun \\
     & $\log N_{el}/N_{tot}$ & $\Delta\log N_{el}/N_{tot}(7300-7815)$ & $\log(N/N_\mathrm{tot})$\\
\hline
C {\sc i}    & $-3.78\pm0.10$ & $-0.05$ & $-3.65$ \\
Na {\sc i}   & $-5.22\pm0.19$ &  0.0 &  $-5.87$ \\
Mg {\sc i}   & $-4.27\pm0.03$ &  0.04 &  $-4.51$  \\
Mg {\sc ii}  & $-4.18$ (:)    &  0.28  & $-4.51$ \\
Si {\sc i}   & $-4.29\pm0.16$ & $-0.22$ & $-4.53$\\
Si {\sc ii}  & $-4.23$ (:)    & $0.13$ &  $-4.53$\\
Ca {\sc i}   & $-5.69\pm0.24$ & $-0.02$ & $-5.73$  \\
Ca {\sc ii}  & $-5.68$ (:)    & $-0.10$ & $-5.73$ \\
Sc {\sc ii}  & $-8.74\pm0.10$ & $-0.34$ & $-8.99$\\
Ti {\sc i}   & $-6.26\pm0.37$ & $-0.10$ & $-7.14 $\\
Ti {\sc ii}  & $-6.69\pm0.29$ & $-0.25$ & $-7.14$ \\
Cr {\sc i}   & $-5.97\pm0.07$ & $-0.15$ & $-6.40$ \\
Cr {\sc ii}  & $-5.88\pm0.09$ & $-0.21$ & $-6.40$\\
Fe {\sc i}   & $-4.09\pm0.15$ & $-0.20$ & $-4.59$\\
Fe {\sc ii}  & $-4.07\pm0.10$ & $-0.15$& $-4.59$ \\
Ni {\sc i}   & $-5.23\pm0.18$ & $-0.18$ & $-5.81$\\
Y {\sc ii}   & $-9.47\pm0.09$ & $-0.16$ & $-9.83$ \\
Ba {\sc ii}  & $-9.00$        & 0.0 & $-9.87$ \\
\hline
\end{tabular}
\label{abundtab}
\end{center}
\end{table}

 By the means of the high-resolution spectroscopic observations we found that the effective temperature, surface gravity  and rotational velocity  of HD\,207561 are 7300$\pm$250 K, 3.7$\pm$0.1 dex and 74$\pm$5 km\,$s^{-1}$, respectively. The abundance analysis shows that HD\,207561 has small under abundances of the Ca and Sc and mild over-abundances of iron peak elements that are the characteristic of the Am stars.
It can be seen from Table \ref{abundtab} that C, O, Ca and Sc show a mild deficiency while all other elements have excess of abundances. Such elemental abundances are generally found in Am stars. Hence we conclude that HD\,207561 is most probably an Am star with mild peculiarities.

 The spectro-polarimetric analysis indicates that, within the observational error of 100 G, HD\,207561 is a non-magnetic star; another support for the Am classification. The location of HD\,207561 in the H-R diagram shows  that the star is slightly evolved from the main sequence and positioned well within the $\delta$-Scuti instability strip. Hence, as an Am star, we would not expect HD\,207561 to exhibit high-overtone pulsation like the roAp stars (and we therefore conclude that the 2000 detection was spurious), 
but it might exhibit low-amplitude $\delta$-Scuti pulsations. Our photometric observations were optimised for detecting rapid oscillations, so we are not able to comment on the latter possibility with the data in hand. It would be interesting to further monitor HD\,207561 for such oscillations in future.

\section*{Acknowledgments}

The authors are grateful to reviewer Dr. Barry Smalley for useful comments and suggestions which led to 
significant improvements in the manuscript. Resources provided by the electronic data bases (VALD, SIMBAD, NASA's ADS) are acknowledged. SJ and PM acknowledge the grant received under the Indo-South Africa Science and Technology Cooperation INT/SAFR/P-3(3)/2009 funded by Departments of Science and Technology(DST) of the Indian and South African Governments. ES is thankful to the Federal programme ``Scientific and educational cadre of innovating Russia 2009--2013'' operated by the Ministry of Education and Science of Russian Federation.  Part of this work was done under the Integrated Long Term Programme (ILTP) supported by the  DST, Govt. of India and Russian Academy of Science vide project INT/ILTP/B-3.16.

\end{document}